\documentclass[iop]{emulateapj}

\shorttitle{The orbital period of  IGR~J16283--4838}
\shortauthors{Cusumano et al.}

\begin{document}
  \title{Swift Observations of the High-mass X-ray Binary IGR J16283--4838 unveil 
  a 288-day Orbital Period}
   \author{G.\ Cusumano\altaffilmark{1},  A.\ Segreto\altaffilmark{1}, 
   V.\ La Parola\altaffilmark{1},
 A.\ D'A\`i\altaffilmark{2}, N.\ Masetti\altaffilmark{3}, G.\ Tagliaferri\altaffilmark{4}}

   \email{G. Cusumano, cusumano@ifc.inaf.it}
\altaffiltext{1}{INAF - Istituto di Astrofisica Spaziale e Fisica Cosmica di Palermo,
        Via U.\ La Malfa 153, 90146 Palermo, Italy}
\altaffiltext{2}{Dipartimento di Fisica, Universit\`a di Palermo, via 
 Archirafi 36, 90123, Palermo, Italy}
\altaffiltext{3}{INAF - Istituto di Astrofisica Spaziale e Fisica Cosmica di Bologna,
via Gobetti 101, 40129, Bologna, Italy}
\altaffiltext{4}{INAF - Brera Astronomical Observatory, via Bianchi 46, 23807, Merate
(LC), Italy }

\begin{abstract}
We report on the temporal and spectral properties of the HMXB
IGR~J16283--4838 in the hard X-ray band.
{\bf We searched the first 88 months of Swift BAT survey
data for long-term periodic modulations.}
We also investigated the broad band (0.2--150 keV) spectral properties of
IGR~J16283--4838 complementing the BAT dataset with 
the soft X-ray data from
the available Swift-XRT pointed observations.
The BAT light curve of IGR~J16283--4838 revealed a periodic modulation at
P$_o=287.6\pm1.7$ days (with a significance higher than 4 standard deviations). The
profile
of the light curve folded at P$_o$ shows a sharp peak lasting $\sim 12$~d,
over a flat plateau. The long-term light curve shows also a $\sim300$~d interval of prolonged 
enhanced emission. The observed phenomenology is {\bf suggestive of a Be nature} of 
IGR~J16283--4838, where the narrow periodic peaks and the $\sim300$~d outburst can be 
interpreted as Type I and Type II outbursts, respectively.
The broad band 0.2--150 keV spectrum can be described with an absorbed
power-law and a steepening in the BAT energy range.
\end{abstract}
\keywords{X-rays: general ---  X-rays: binaries ---
X-rays: individuals (IGR J16283-4838) }

\section{Introduction}\label{intro} 
High-mass X-ray binaries (HMXB) consist of  a magnetized neutron star
or a black hole, and a massive OB star.
Three different sub-classes are distinguished,  depending on {\bf the
mechanism} that transfers matter from the massive star onto the compact object:
(i) through Roche lobe overflow via the formation of an accretion disc around the compact
object, (ii) capturing the high velocity stellar wind of an early type star
or (iii) capturing material from the extended envelope of the equatorial 
circumstellar disk of a main sequence Be star.
The majority of the HMXBs observed up to now are in systems with a Be star
as a companion:  the low velocities and the high densities present in the Be
envelope provide a dense environment that favour accretion
\citep{bondi44,negueruela07}. 
Be/X-ray binaries are often observed as transient systems and display 
two types of X-ray outbursts. Type I outbursts are periodic and occur
close to the periastron passage of the neutron star in an eccentric orbit, 
when the accretion is enhanced due to the mass capture from the equatorial disk of
the Be star. 
Type II outbursts, usually much brighter than Type Is, are not
related to the orbital phase. They are caused by  sudden disk instabilities 
that produce a prolonged accretion of a huge fraction of the Be star disk, 
and can be sustained for many orbital periods.

The  Burst Alert Telescope (BAT, \citealp{bat}) on board the Swift
observatory \citep{swift} has been scanning the sky since December 2004.
After  88 months the survey has achieved a sensitivity down to some units of
$\rm 10^{-12} erg~cm^{-2} s^{-1}$ in the 15--150 keV band and, thanks to the
large field of view (1.4 steradians, half coded) and to the Swift pointing strategy
that maximizes the observing duty cycle of the spacecraft instruments, BAT
covers a large fraction of the sky per day (50\%--80\%),
recording timing and spectral information for any detected source.

\begin{figure*}
\begin{center}
\centerline{\includegraphics[width=18.cm,angle=0]{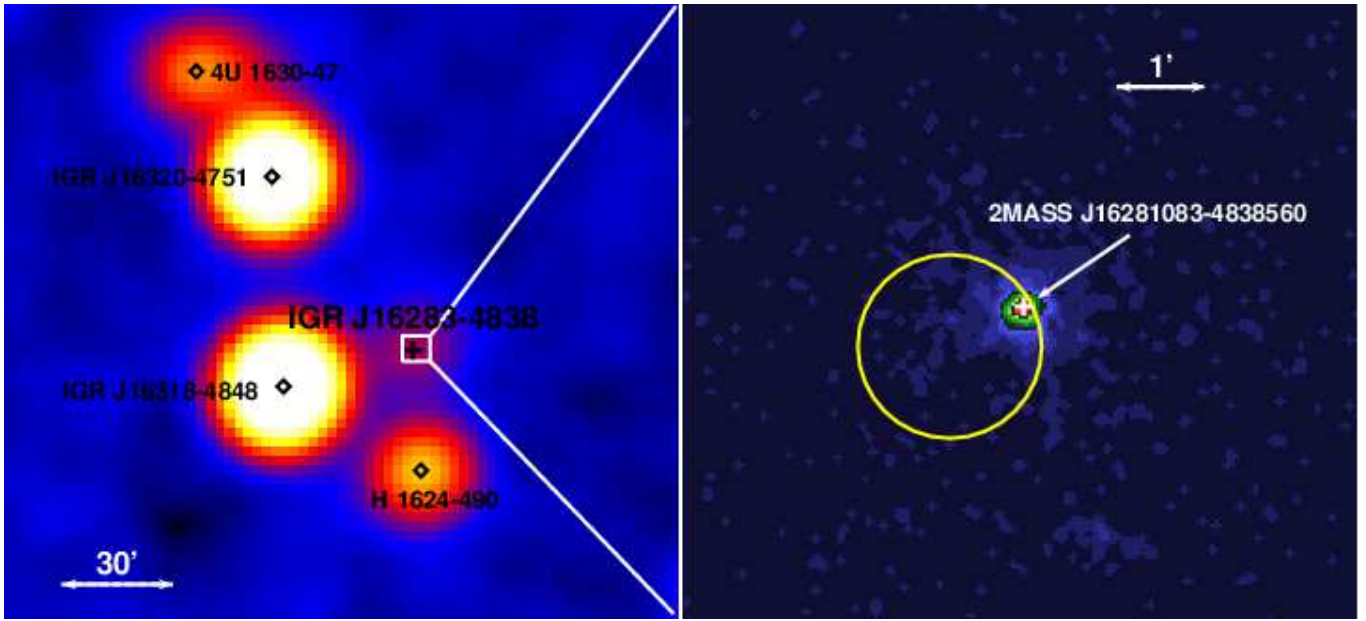}}
\caption[IGR~J16283--4838 sky maps]{ 
Left panel:  BAT significance map in the sky region around IGR~J16283--4838.
Right panel: 0.2-10 keV XRT image with superimposed the position of the NIR
counterpart 2MASS~J16281083--4838560, marked with a cross, and the BAT 99\% 
error circle of 1.02 arcmin radius (yellow circle). 
                }
                \label{map} 
        \end{center}
        \end{figure*}

We analyze the soft and hard X-ray data collected by Swift on 
IGR~J16283--4838. This source was discovered by INTEGRAL as a hard X-ray 
transient during a core
program observation of the Norma Arm \citep{atel456}. A Swift-XRT
observation constrained better its position, setting it at RA= 16h 28m 10.7s, 
Dec=--48deg 38' 55'' (J2000), with a 5'' radius uncertainty \citep{atel459}. The source 
was observed to be variable, with its flux rising by  a factor of 2 on a time-scale 
of a few days \citep{atel458,atel465}.
The X-ray spectrum of the source showed a strong and variable intrinsic absorption 
($\rm N_H = 0.4 - 1.7 \times 10^{23}~cm^{-2}$). {\bf The continuum could be adequately 
modeled by either a power law with} $\Gamma=1.12$ or with a black body with a temperature 
kT=2.0 keV \citep{beckmann05}.
\citet{atel460} indicated the near-infrared (NIR) source 2MASS~J16281083--4838560 
(RA = 16h 28m 10.83s, Dec =--48deg 38' 56.1", J2000, 1.7" from the XRT position, with 
magnitudes K$=13.95\pm0.06$, J $>$ 16.8, and H $>$ 15.8) 
as the possible counterpart of IGR~J16283--4838. 
The Galactic Legacy Infrared Midplane Survey Extraordinaire
(GLIMPSE, \citealp{benjamin03}) data show a source positionally consistent with the 2MASS source.
However, a K band image acquired at the 6.5m Magellan-Baade telescope revealed that
2MASS~J16281083--4838560 is indeed a blend of point sources, where the brightest resolved one 
has a K band magnitude of $\sim14.1$ \citep{atel478}. 
No UV or optical source was found within the XRT error box either in Swift-UVOT images, 
nor in the Digitized Sky Survey \citep{beckmann05}.
\citet{pellizza11}, using optical and NIR images, spectra and polarimetry, 
suggest that the brightest object of the 2MASS~J16281083--4838560 blend is the counterpart 
to IGR~J16283--4838. The source
shows NIR spectral features of late-O or early-B supergiant stars, 
strengthening the hypothesis that IGR~J16283--4838 is a highly absorbed HMXB. 

In this Letter, section~\ref{data} describes the Swift data reduction, 
section~\ref{timing} reports on the timing analysis, 
section~\ref{spectral} describes the broad band spectral analysis; in 
section~\ref{conclusion} we briefly discuss our results. 

\section{Observations and data reduction}\label{data}

The BAT survey raw data of the first 88 months of the Swift
mission were processed {\bf with dedicated} software \citep{segreto10} 
that computes all-sky maps in  several energy bands between 15 and 150 keV,
performs source detection on {\bf these maps, and, for each detected source,
produces} light curves and spectra.
IGR~J16283--4838 was detected with highest significance in the 15--45 keV all-sky map 
with a signal to noise ratio of 21 standard deviations. 
Fig.\ref{map} shows the BAT image of the sky region around 
IGR~J16283--4838.  The source was inside the BAT field of view 
for a total of 30.5 Ms. The BAT timing analysis is performed on 
the  15--45 keV light curve extracted with the maximum available time 
resolution ($\sim300$ s). The spectrum was analyzed 
using the official BAT spectral redistribution matrix
\footnote{http://heasarc.gsfc.nasa.gov/docs/heasarc/caldb/data/swift/bat/index.html}.

{\bf The Swift-XRT} \citep{xrt} observed IGR~J16283--4838  {\bf on} 2005 April 13 
(Obs ID 00067133001)  and April 15 (Obs ID 00067133002). The source was observed 
both in Photon Counting (PC) mode and in {\bf Windowed} Timing (WT) mode \citep{hill04}
in each observation. 
The details on the two observations are reported in Table~\ref{log}.
The XRT data were processed with standard procedures ({\sc xrtpipeline} 
v 0.12.4) using {\sc ftools} in the {\sc heasoft} package (v 6.8) and the 
products were extracted {\bf with} grade filtering of 0-12 and 0-2 for PC 
and WT data, respectively. The source events for data collected in PC mode 
were extracted from a circular region of 20 pixel radius (1 pixel = 2.36'') 
centered on the source position as determined with {\sc xrtcentroid}; 
the background for spectral analysis was extracted from an annular region 
centered on the source, with an inner radius of 70 pixels and an outer radius of
130 pixels.  Source events in WT mode were extracted from a 20 pixel wide portion of the WT strip 
centered on the source position; the background was selected from a 
region sufficiently 
far ($>2$ arcmin) to avoid  the contamination due to the PSF tail 
of IGR~J16283--4838. All source event arrival times were converted to 
the {\bf Solar System barycenter} with the task 
{\sc barycorr}\footnote{http://http://heasarc.gsfc.nasa.gov/ftools/caldb/help/barycorr.html}.
XRT ancillary response files were generated 
with {\sc xrtmkarf}\footnote{http://heasarc.gsfc.nasa.gov/ftools/caldb/help/xrtmkarf.html}. 
Spectra relevant to the same observing mode were summed to obtain a
single {\bf source spectrum and a single background spectrum}. The ancillary files were 
combined using {\sc addarf} weighting them by the exposure times of the 
relevant source spectra. Each spectrum was re-binned with a 
minimum of 20 counts per energy channel, in order to allow the use of the 
$\chi^2$ statistics. We used the spectral redistribution matrix v013 and the 
spectral analysis was performed using {\sc xspec} v.12.5.  
{\bf Uncertainties are stated at} 90\,\% confidence level for a single parameter, if not stated otherwise.

\begin{table*}
\caption{XRT observations log \label{log}} 
\scriptsize
\begin{center}
\begin{tabular}{r l l l l l l l l} \tableline
Obs \# & Obs ID & $T_{start}$ & $T_{elapsed}$ & PC Exposure & WT Exposure & PC Rate & WT Rate & Orb. Phase \\
       &        &             &  (s)        & (s)         & (s)   & (c/s)& (c/s) &  \\ \tableline \tableline
1      &00067133001 & 53473.58 & 2422 & 647 & 812   &$0.28\pm0.02$ &$0.39\pm0.02$  &0.984  \\
2      &00067133002 & 53475.01 & 81644& 5116& 1139.9&$0.121\pm0.005$ &$0.21\pm0.01$&0.979  \\ \tableline
\end{tabular}
\end{center}
\tablecomments{The quoted orbital phase refers to the profile reported
in Figure~\ref{period}b.}
\end{table*}

\begin{figure}[h]
\centerline{\includegraphics[width=7.7cm]{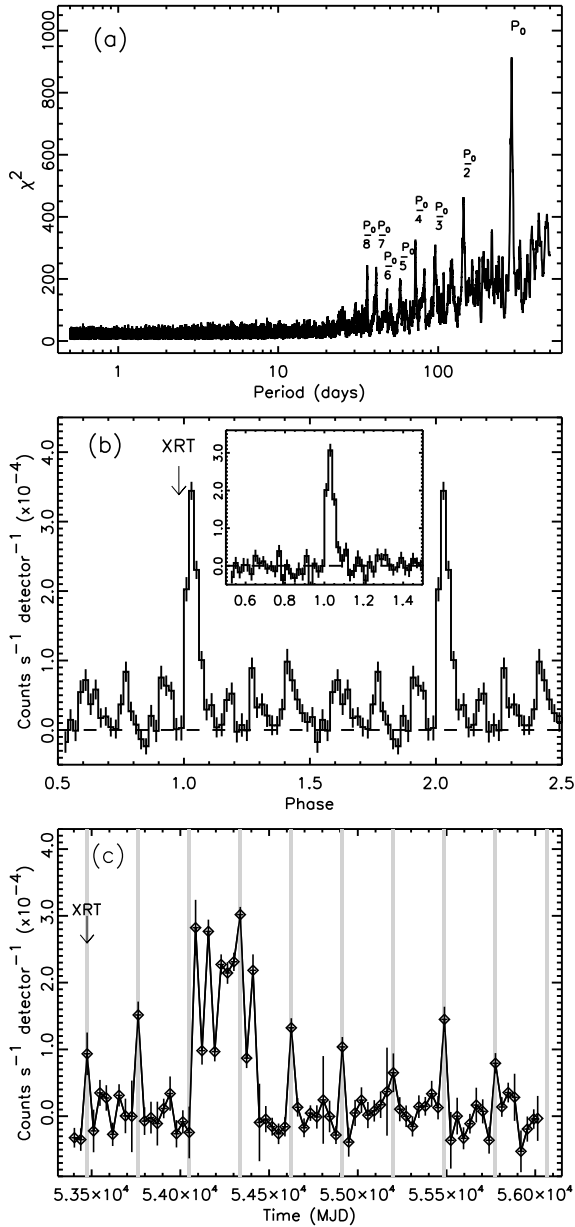}}

\caption[]{{\bf (a)}: Periodogram of {\it Swift}-BAT (15--45\,keV) data for
IGR\,J16283--4838.
{\bf (b)}: {\it Swift}-BAT light curve folded at a period $P=287.6\pm1.7$\,d,
with 50 phase bins. The arrow points to the phase  of 
the two XRT observations. The inset shows the folded profile obtained excluding 
the BAT data between MJD 54080 to 54420 characterized by a prolonged enhanced emission.
{\bf (c)}: BAT light curve. The bin length corresponds to a time interval of  
$P_o/8=35.95$\,days.  The vertical shaded areas, each having a width of 12 days,  
are in phase with the peak in the folded profile.
                
                \label{period}
}
\end{figure}

\section{Timing analysis and results}\label{timing} 

We  produced the periodogram {\bf from} the long term Swift-BAT 15--45 keV light curve 
{\bf by} applying  a folding technique \citep{leahy83} and searching in the 0.5--500 d 
period range with a period spacing given by P$^{2}/(N  \,\Delta$T$_{\rm BAT})$ (P is 
the trial period, $N=16$ is the number of phase bins used to build the profile 
and $\Delta$T$_{\rm BAT}\sim 230.6$ Ms is the  data span length). 
The  average rate in  each profile phase bin was evaluated by {\bf weighting 
each rate with the inverse square of its statistical} error. 
This procedure is appropriate as the data are characterized by a large span of 
statistical errors, being background dominated, and collected, for each source, 
over a wide range of off-axis directions.
The periodogram (Fig.~\ref{period}a)  shows the 
presence of several prominent features. The highest one is 
at P$_o=287.6\pm1.7$ d ($\chi^2\sim 930$), where the centroid and the 
error are {\bf taken to be the
location of the peak and the standard deviation obtained from a
Gaussian fit} to the $\chi^2$ feature at P$_o$.  
Other less significant features are clearly visible in  the
periodogram and they correspond to a series of sub-multiples of P$_o$.  

Figure~\ref{period}b shows the intensity profile folded at  P$_o$  
(T$_{\rm epoch}=54619.5 $ MJD) using 50 phase bins. The profile is characterized by a
prominent narrow peak and several lower peaks.
The main peak centroid and its width have been evaluated with a Gaussian fit, obtaining
 a centroid phase at $0.033\pm 0.002$ (that corresponds to  
MJD $(54628.5\pm 0.6)\pm  n \times $P$_o$) and a standard deviation of $0.021\pm0.002$
($6.0\pm0.6$ days).

The profile shape explains the presence of 
the large number of P$_o$ sub-multiples in the periodogram: folding the
light curve with a
period equal to P$_o/2$, the peak is added coherently to the profile every two cycles, 
producing a sub-harmonic in the periodogram. The sharper and shorter the profile peak is, the
more sub-multiples will appear in the periodogram. We verified 
this expectation by simulating a 88-month BAT light curve with a squared periodic
(287.6 days) peak lasting 12 days superimposed over a white noise signal with an average 
intensity a factor of ten lower than the peak intensity. The periodogram obtained by the simulated light
curve satisfactorily reproduced the periodogram obtained from the real data.

 Figure~\ref{period}c shows the 88-month BAT light curve with a time bin {\bf length} of 
P$_o/8= 35.95$\,days. The vertical shaded bars mark the times corresponding to the main peak
phase: the periodic enhancement of the hard X-ray intensity is clearly 
visible. The source shows also a prolonged time interval of high intensity starting 
at the beginning of 2007 and lasting $\sim 300$ days.
We have verified that this is the cause for the presence of
the lower peaks in the folded profile: if we repeat the folding excluding this time interval
(MJD 54080 to 54420) the resulting folded profile shows the main peak over a flat
plateau consistent with zero intensity (inset in Figure~\ref{period}b).

The significance of  the  feature at P$_o$  cannot  be evaluated  {\bf using 
$\chi^2$ statistics}  because the long term  variability of 
IGR~J16283$-$4838  causes the $\chi^2$ distribution in the periodogram
to {\bf strongly deviate in terms of both average and
fluctuation amplitude from what is expected for white noise} 
(average $\chi^2=N-1$).
The  significance  of  P$_o$  is therefore evaluated as follows:

\begin{enumerate}
\item We fitted the periodogram with a  second order polynomial and
derived the average trend of the $\chi^2$ versus P and the spread
of the  $\chi^2$ values around the best fitting parabola.

\item We built a set of n1=10000 fake light curves, produced by splitting
the IGR~J16283--4838 data into
portions of n2=900 bins and performing a random swap of the corresponding
rate blocks; this was repeated n3=3 times for each fake curve, using a
different starting point for the sectioning. The values of n2 and n3 were chosen after
testing over several combinations of values which one reproduced better the $\chi^2$
distribution of the observed periodogram both in average trend and in scatter.


\item {\bf The folding analysis was performed for each fake light curve to
obtain a periodogram covering the 0.5--500 day period range. The
highest $\chi^2$ value in the entire period range in the set of 10000
periodograms was 794. The probability that a random fluctuation would
produce a $\chi^2$ value as high as 930 in one particular periodogram is
therefore lower than  $\rm 1/n1=1\times 10^{-4}$ and corresponds to a
significance higher than $\sim 4$ standard deviations.}

\end{enumerate}

We performed a timing analysis on the {\bf XRT data in order to search for} 
the presence of a periodic modulation.
The arrival times of the events in PC mode were randomized within the XRT-PC time resolution bin 
($\delta T_{XRT}$=2.5073 s) to avoid systematics caused by the read-out time.
Moreover, XRT observations are fragmented into snapshots  of different duration and
time separation. This causes the presence of spurious features in the timing analysis 
that could affect the detection of true periodic signals.
To avoid these systematics we performed a folding analysis on each snapshot with 
an exposure time higher than 500 s, searching in a period range [$\delta
T_{XRT}$~:~100] s and [1~:~100] s, for PC and WT mode, respectively. 
The periodograms obtained from snapshots belonging to the same observation
and observing mode were summed together. We did not find any significant feature 
in the resulting periodograms.


\section{Spectral Analysis} \label{spectral}

Before performing {\bf a broad-band} spectral analysis of non-simultaneous data 
we verified that no significant spectral {\bf variability is present in either of 
the two XRT observations or during the BAT monitoring}.
{\bf The two XRT spectra were each fit with an
absorbed power-law spectral model. The results for the two spectra
were consistent with a single set of parameters.}
To check if spectral variability is present during the BAT monitoring we 
produced three BAT spectra. {\bf Spectrum 1 was made from data obtained during} the 
2007 outburst (Figure~\ref{period}c). 
{\bf Spectra 2 and 3 were constructed excluding data
taken during the outburst. Spectrum 2 was built using data taken
during phases 0.0-0.06 in Figure~\ref{period}b, i.e., phases corresponding 
to the peak, while Spectrum 3 was built from data taken in the off-peak phase
range 0.06-1.0. Spectra 1 and 2 were independently fit with a power
law; the best fit photon indices are mutually consistent within the
uncertainties. }
In Spectrum 3 the source is not detected, with an upper limit on the count rate of $8.5\times
10^{-6}$ counts s$^{-1}$ detector$^{-1}$.

These intermediate results {\bf support} the assumption of no significant spectral
evolution {\bf in either soft or hard X-rays}. 
{\bf The broad band spectral analysis was then performed
on Spectrum 1 together with the XRT spectra 
(hereafter XB1). The trial models for the XRT spectra were multiplied
by constants (XRT-PC, C$_{PC}$; XRT-WT, C$_{WT}$) to account for different
flux levels and for inter-calibration uncertainties. A similar broad
band analysis was performed on Spectrum 2 together with the XRT spectra 
(hereafter XB2).}

{\bf The best fit of the absorbed power law model
({\tt cons*phabs*powerlaw}) to XB1 yielded a $\chi^2$ of 214 for 75 degrees of
freedom (d.o.f.). Therefore XB1 is not well-described by this model.
We obtained a marginally acceptable fit for XB2($\chi^2 = 110.5$, 75
d.o.f.).}
We have then modeled the two spectra with a power-law modified by a cutoff 
({\tt cons*phabs*cutoffpl}), obtaining in both cases a good fit of the data 
($\chi^2$ of 89.5 with 74 d.o.f for XB1 and $\chi^2$ of 85.2 with 74 d.o.f for XB2), with
consistent spectral parameters (reported in Table~\ref{fit}).
Assuming a similar spectral shape for Spectrum 3 and using the same spectral parameters 
obtained for XB2 we obtain a $3\sigma$ upper limit on the 0.2--150
keV flux of $\rm 2.9\times 10^{-12}~erg~cm^{-2}~s^{-1}$.

\begin{figure*}
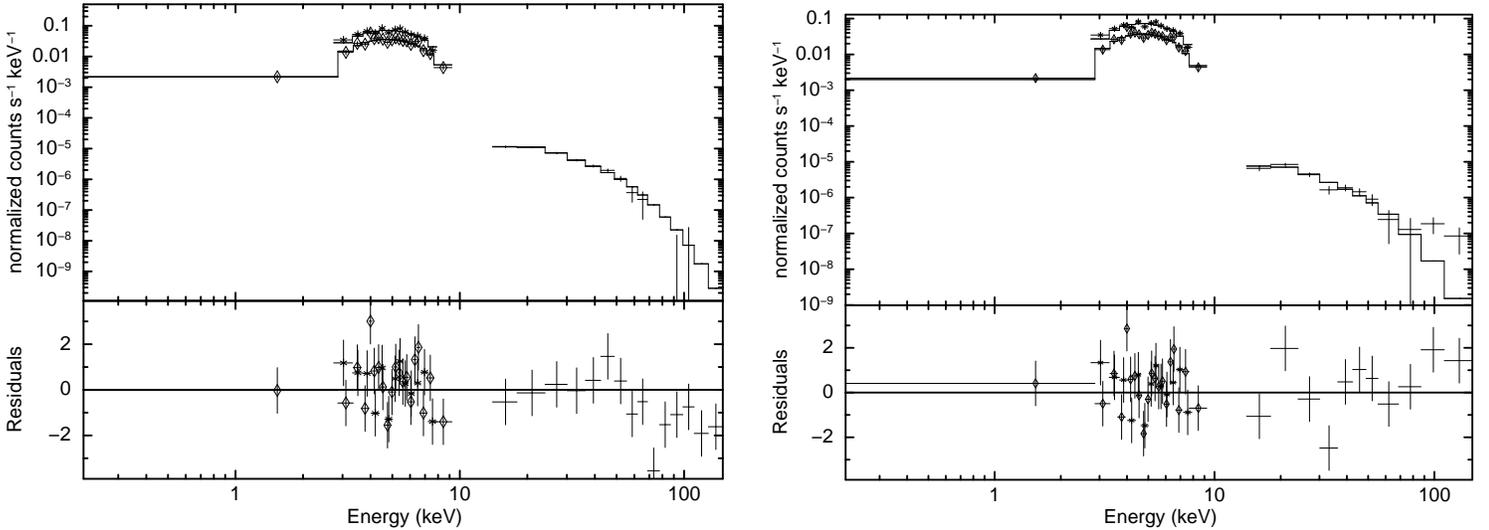

\begin{center}
\centerline{\includegraphics[width=7cm,angle=270]{f5a.ps}
            \includegraphics[width=7cm,angle=270]{f5b.ps}}
\caption[]
{\bf IGR~J16283--4838 broad band spectra (data, best fit model and 
{\bf residuals in units of $\sigma$}) fitted with a 
{\tt phabs*(cutoffpl)} model (WT data are marked with diamonds, PC data are marked with stars). 
{\bf Left panel}. XB1: the XRT spectra are combined with the BAT spectrum 
constructed from data taken during the 2007 outburst. {\bf Right panel}. 
XB2: the XRT spectra are combined with the BAT spectrum 
constructed from data taken during the periodic
peaks. 
  }
                \label{spec}
        \end{center}
        \end{figure*}

Figure~\ref{spec} shows the data, best fit model ({\tt cons*phabs*cutoffpl}) and 
residuals for XB1 (left panel), and for XB2 (right panel).

\begin{table}
\caption{Best fit spectral parameters.  \label{fit}}
\tabletypesize{\scriptsize}
\renewcommand\arraystretch{1.3}
\renewcommand\tabcolsep{3pt}
\begin{tabular}{ r l l l}
\tableline
Parameter        & Powerlaw                            & Cutoff powerlaw                  & Units    \\ \tableline
\multicolumn{4}{c}{XRT + BAT Outburst Spectrum (XB1)}\\ \tableline
N$_{\textrm{H}}$ &		 & $13^{+2}_{-2}\times 10^{22} $  & $\rm cm^{-2}$\\
$\Gamma$         &		 &$0.37^{+0.35}_{-0.35}$		  &	     \\
$E_{cut}$        &		 &$14^{+3.0}_{-2}$		       & keV\\
$N$              &		 &$ 1.3^{+1.5}_{-0.8}\times 10^{-3}$	&$\rm ph/(keV~cm^{2}s)$ @1 keV \\
$\rm C_{PC}$     &		 & $1.4^{+1.7}_{-0.8}$  		&\\
$\rm C_{WT}$     &		 & $2.6  ^{+0.8}_{-0.6}$		&\\
$\rm F_{0.2-150}$ &		 & $1.4\times 10^{-10}$&$\rm erg~cm^{-2} s^{-1}$ \\
$\chi^2$         &		 &89.5 (74 dof) 			&\\ 
\tableline
\multicolumn{4}{c}{XRT + BAT Peak Spectrum (XB2)}\\ \tableline
N$_{\textrm{H}}$ &$23^{+2}_{-2}\times 10^{22}$   &$15^{+3}_{-2}\times 10^{22} $  & $\rm cm^{-2}$\\
$\Gamma$         &$2.36^{+0.19}_{-0.18}$         &$0.8^{+0.5}_{-0.5}$		       &	  \\
$E_{cut}$        &                               &$17^{+10.0}_{-5}$			& keV\\
$N$              &$ 6^{+6}_{-3}\times 10^{-2}$   &$ 2^{+5}_{-1}\times 10^{-3}$    &$\rm ph/(keV~cm^{2}s)$ @1 keV \\
$\rm C_{PC}$     &$0.8^{+0.3}_{-0.2}$            & $1.7^{+0.7}_{-0.5}$  		&\\
$\rm C_{WT}$     &$1.4  ^{+0.5}_{-0.4}$          & $3.2  ^{+1.4}_{-1.0}$		&\\
$\rm F_{0.2-150}$ &$4.7\times 10^{-10}$           & $1.05\times 10^{-10}$&$\rm erg~cm^{-2} s^{-1}$ \\
$\chi^2$         &110.5 (75 dof)                 &85.2 (74 dof) 			&\\ \tableline
\end{tabular}
\renewcommand\arraystretch{1}
\tablecomments{$\rm C_{PC}$ and $\rm C_{WT}$ are 
are {\bf constant multiplicative factors applied to the model}
in order to match the XRT PC 
and WT spectra to the BAT spectra.  We report
unabsorbed fluxes in the 0.2--150 keV band.}
\end{table}

\section{Discussion and conclusions}\label{conclusion}
The analysis of the BAT survey data of the HMXB IGR~J16283--4838 
unveiled a modulation of its hard X-ray emission corresponding to a period
of P$_o=287.6\pm 1.7$ days. This modulation, whose significance 
is higher than 4 standard deviations, is clearly visible in the BAT light curve
(Fig.~\ref{period}c) and can be interpreted as the orbital period of the binary system. 
Such long orbital periods are typical of Be/X-ray binaries (e.g \citealp{corbet86}).
Moreover, we found that the source has undergone a long 
period of strong activity between  December 2006 and November 2007, indicating enhanced
accretion for more than one entire orbit.
The BAT light curve {\bf outside} the prolonged outburst, folded at P$_o$ is characterized by a 
a narrow peak over a flat plateau whose intensity is consistent with zero (inset in 
Figure~\ref{period}b). The peak  duration is $\sim 12$ days. 
Thus a large fraction of the emission of this source is concentrated in a short fraction of its
orbit. This behavior is often
observed in X-ray binaries with a Be star as a companion, and it is
explained {\bf as the consequence of the
enhancement of accretion when the neutron star is at or near the
periastron of a highly eccentric orbit. These are the Type I
outbursts.  The $\sim 300$ day long outburst} can be
interpreted as a Type II outburst, a
further signature of the Be nature of the system.
The broad band 0.2--150 keV spectral analysis shows that the spectra selected during the 
outburst and during the periodic peaks can 
be both modeled with an absorbed powerlaw modified by a high energy cutoff, with consistent best
fit parameters. Assuming a distance of  13.6--21.6 kpc \citep{pellizza11} and
isotropic emission,
the 0.2--150 keV luminosities are $\rm (5.2\pm 2.3)\times10^{36}erg s^{-1}$ and 
$\rm (3.9\pm 1.8)\times10^{36}erg s^{-1}$, respectively. During the plateau phase the source 
is not detected, with a $3\sigma$ upper limit on 
the 0.2--150 keV luminosity of $\rm 2.1\times10^{35}erg s^{-1}$.

\citet{pellizza11} identified the NIR counterpart of IGR~J16283--4838 {\bf as} the brightest object of the
blend of three sources catalogued as  2MASS~J16281083--4838560 and classified it as a late-O or early-B supergiant.
Under this assumption, and assuming a typical
mass $\rm M_{\star}\sim 25 M_{\odot}$ and a typical radius 
$\rm R_{\star}\sim 30 R_{\odot}$ \citep{lang92} for a B0 star, we can infer the length of 
the orbital semi-major axis  through the third Kepler's law:
\begin{equation}
a=(G P_o^2~(M_{\star}+M_{\rm X})/4\pi^2)^{1/3} \simeq 546.1 R_{\odot} = 18 R_{\star}
\end{equation}
where we assumed the compact object to be a neutron star with 
$M_{\rm X}=1.4 M_{\odot}$.
Such a wide orbital separation is not common in systems with a supergiant 
companion {\bf star and wind-driven} accretion. 
Assuming the above referenced values of radius and mass for the companion star, the 
expected accretion rate would be \citep{frankkingraine}:

\begin{equation}
\dot{M} = - \dot{M_w} \left( \frac{M_{NS}}{M_{\star}} \right)^2 \left( \frac{R_{\star}}{a} \right)^2 
\end{equation}

Even considering an extreme  wind loss rate ($M_w$) of $10^{-5}$  M$_{\sun}$ year$^{-1}$ and 
an accretion efficiency of 10\%, the expected luminosity would correspond to $\sim$  10$^{35}$
erg s$^{-1}$, that is an order of magnitude lower than the 
observed values. It is therefore quite unlikely that the system harbors a O 
supergiant with a wind-fed NS. 

An alternative possibility is that the counterpart of IGR~J16283--4838 is one of the other two
sources that compose 2MASS~J16281083--4838560. Both of them are fainter than the one chosen by
\citet{pellizza11} as the counterpart, but closer to the position of the soft X-ray
counterpart observed with XRT. Unfortunately, we have no information on 
the spectral type {\bf of either of these} two objects. 

We have performed timing analysis on the XRT data, searching for periodicities in the range
1--100~s, finding no evidence for periodic modulations.  
If the system were a Be/X-ray binary, the Corbet correlation between spin period and orbital 
period \citep{corbet86} would suggest a spin period around $1000$
s, that is beyond the range of periodicities that can be revealed with the
available XRT data. Indeed, such long periodicities are not easily
revealed in XRT observations that, because of the low satellite orbit, 
are split into many snapshots (one per orbit, where each orbit lasts $\sim 5800$ s)
lasting only a fraction of the orbit itself (between $\sim 500$ and $\sim1000$ s for this
source). The timing analysis suffers from such fragmentation as it induces
strong systematics that can hide the presence of long spin periods.

\begin{acknowledgements}
We thank the anonymous referee for his helpful comments and suggestion. 
This work has been supported by ASI grant I/011/07/0.
\end{acknowledgements}

\bibliographystyle{aa}

\begin{thebibliography}{}


\bibitem[Barthelmy et al.(2005)]{bat} Barthelmy, S.~D., et 
al.\ 2005, Space Science Reviews, 120, 143 

\bibitem[Beckmann et al.(2005)]{beckmann05} Beckmann, V., Kennea, 
J.~A., Markwardt, C., et al.\ 2005, \apj, 631, 506 

\bibitem[Benjamin et al.(2003)]{benjamin03} Benjamin, R.~A., 
Churchwell, E., Babler, B.~L., et al.\ 2003, \pasp, 115, 953 

\bibitem[Bondi \& Hoyle(1944)]{bondi44} Bondi, H., \& Hoyle, F.\ 1944, \mnras, 
104, 273 


\bibitem[Burrows et al.(2004)]{xrt} Burrows, D.~N., Hill, 
J.~E., Nousek, J.~A., et al.\ 2004, \procspie, 5165, 201 


\bibitem[Corbet(1986)]{corbet86} Corbet, R.~H.~D., 1986, \mnras, 220, 1047

\bibitem[Dickey \& Lockman(1990)]{dickey90} Dickey, J.~M., \& Lockman, F.~J.\ 
1990, \araa, 28, 215 

\bibitem[Frank et al.(2002)]{frankkingraine} Frank, J., King, A., 
\& Raine, D.~J.\ 2002, Accretion Power in Astrophysics, by Juhan Frank and Andrew King and Derek Raine, pp.~398.~ISBN 0521620538.~Cambridge, UK: Cambridge University Press, February 2002.,  

\bibitem[Gehrels et al.(2004)]{swift} Gehrels, N., et al.\ 
2004, \apj, 611, 1005 

\bibitem[Hill et al.(2004)]{hill04} Hill, J.~E., Burrows, 
D.~N., Nousek, J.~A., et al.\ 2004, \procspie, 5165, 217 

\bibitem[Kalberla et al.(2005)]{kalberla05} Kalberla, P.~M.~W., Burton, 
W.~B., Hartmann, D., et al.\ 2005, \aap, 440, 775 

\bibitem[Kennea et al.(2005)]{atel459} Kennea J. A., Burrows D. N., Nousek
J. A., Chester M., Roming P., Barthelmy S., Gehrels N., Beckmann V., Soldi S.,
2005, ATEL 459

\bibitem[Lang (1992)]{lang92} Lang, K.~R.\ 1992, Astrophysical
Data I.~ Planets and Stars, X, 937 pp.~33 figs..~ Springer-Verlag Berlin
Heidelberg New York

\bibitem[Leahy et al.(1983)]{leahy83} Leahy, D.~A., Darbro, W.,
Elsner, R.~F., et al.\ 1983, \apj, 266, 160


\bibitem[Markwardt et al.(2005)]{atel465} Markwardt C. B., Swank J. H.,
Smith E.,    2005, ATEL 465

\bibitem[Negueruela et al.(2007)]{negueruela07} Negueruela, I., 
Smith, D.~M., Torrej{\'o}n, J.~M., \& Reig, P.\ 2007, 
ESA Special Publication, 622, 255 

\bibitem[Paizis et al.(2005)]{atel458}  Paizis A., Miller J. M., Soldi S.,
      Mowlavi N., 2005, ATEL 458

\bibitem[Pellizza et al.(2011)]{pellizza11} Pellizza, L.~J., Chaty, S., 
\& Chisari, N.~E.\ 2011, \aap, 526, A15 

\bibitem[Rodriguez \& Paizis(2005)]{atel460}  Rodriguez, J. \& Paizis A., 2005, ATEL 460

\bibitem[Segreto et al.(2010)]{segreto10} Segreto, A., Cusumano, G., Ferrigno, C., La Parola, V., Mangano, V., Mineo, T., \& Romano, P.\ 2010, \aap, 510, A47 

\bibitem[Soldi et al.(2005)]{atel456} Soldi, S., Brandt S., Domingo Garau A.,
     Grebenev S. A., Kuulkers E. , Palumbo G. G.C., Tarana A., 2005,
ATEL 456

\bibitem[Steeghs et al.(2005)]{atel478} Steeghs  D., Torres M. A.P., 
Jonker P. G., Miller J., Green P., Rakowski C., 2005,
ATEL 478


\end{thebibliography}

{}

\end{document}